\documentclass[aps,amsfonts,prl,twocolumn,showpacs]{revtex4-1}

\usepackage{subfigure}
\usepackage{wrapfig}
\usepackage{textcomp}
\usepackage{graphicx}
\usepackage{amssymb}
\usepackage{amsmath}
\usepackage{bm}
\usepackage{amsmath, amsthm, amssymb}

\def\beq{\begin{equation}}
\def\eeq{\end{equation}}
\def\bea{\begin{eqnarray}}
\def\eea{\end{eqnarray}}

\begin{document}

\title{Reply to Comment on ``Quantum quasicrystals of spin-orbit coupled dipolar bosons''}

\author{Sarang Gopalakrishnan$^1$, Ivar Martin$^2$, and Eugene A. Demler$^1$}

\affiliation{$^1$Department of Physics, Harvard University, Cambridge MA 02138, USA \\ 
$^2$ Materials Science Division, Argonne National Laboratory, Argonne, Illinois 60439, USA}

\maketitle

\noindent \textbf{Gopalakrishnan, Martin, and Demler reply:} In a recent Letter~\cite{gmd}, we proposed a scheme for realizing quantum quasicrystals using spin-orbit coupled dipolar bosons. We remarked that these quantum quasicrystals have additional ``phason''-like modes compared with their classical counterparts. A recent comment by Lifshitz~\cite{lifshitz-comment} contests this claim. We argue here that our enumeration of gapless modes is indeed the physically relevant one. Whether the additional modes are \emph{phasons} is, however, a matter of definition. They resemble conventional phasons in being local rearrangements, but can be mathematically distinguished, as discussed below.

The difference between our counting of gapless modes and that of Ref.~\cite{lifshitz-comment} arises because we are counting different things. In our Letter~\cite{gmd}, we enumerated the continuous symmetries of our Hamiltonian [Eq.~(1) of Ref.~\cite{gmd}], and \emph{explicitly} noted that higher-order terms in the Bose fields would break some of these symmetries. Ref.~\cite{lifshitz-comment} enumerates the symmetries that survive to \emph{all} orders in the Bose fields; there are naturally fewer of these. 

The disagreement between our analyses is thus about when it is appropriate to regard a mode as gapless, or, equivalently, when it is appropriate to regard a physical system as possessing a continuous symmetry. Such symmetries are never truly exact in \emph{any} physical system: there always exist symmetry-breaking perturbations such as stray fields, anisotropies, etc. that gap out the putative Goldstone modes. Nevertheless, a system can be regarded as having a continuous symmetry provided that such perturbations are small compared with the relevant physical scales (in our case, the two-body interaction energy and the temperature).

In our Letter, we stated that higher-order terms in the fields can be neglected because they are suppressed at low densities; another way of putting it is that the gaps they generate are negligibly small. We now estimate the relevant energy scales, beginning with the observation that the order parameters in our construction are the microscopic Bose fields themselves. Thus, the coefficient of the $(\phi^\dagger \phi)^2$ term can be related to the microscopic two-body interaction; that of the $(\phi^\dagger \phi)^3$ term can be related to the three-body interaction, etc. For low-temperature s-wave and dipolar scattering, these coefficients can be expressed in terms of the contact and dipolar scattering lengths $a_s$ and $a_d$ respectively; in the parameter regime considered in Ref.~\cite{gmd} the two scales are similar, and are both approximately $100$ Bohr radii, or $5$ nm. As these scattering lengths are much smaller than the confinement scale ($\agt 250$ nm~\cite{gmd}) we can neglect the influence of confinement on scattering properties. Thus, the two-body contact interaction energy scale is~\cite{pethick}

\beq
E_{\mathrm{2-body}} \approx \frac{\hbar^2 \rho_{\mathrm{2D}}}{2m} \frac{a_s}{d_z}
\eeq
where $\rho_{\mathrm{2D}}$ is the density, $m$ is the particle mass, and $d_z$ is the transverse confinement scale. Similarly, the energy density due to the three-body contact interaction is given by~\cite{fedichev}

\beq
E_{\mathrm{3-body}} \approx \frac{\hbar^2 \rho_{\mathrm{2D}}^2 a_s^4}{2m d_z^2}
\eeq
up to a dimensionless parameter of order unity (estimated in Ref.~\cite{fedichev} as $3.9$). Similar formulas hold for the dipolar case if one replaces $a_s$ with $a_d$~\cite{ticknor, levdy}. Thus, the ratio of three-body to two-body interaction scales is given by $\mathrm{max}(\rho_{\mathrm{2D}} a_s^3/d_z, \rho_{\mathrm{2D}} a_d^3/d_z)$. As discussed in Ref.~\cite{gmd}, the greatest achievable densities are $\rho_{\mathrm{2D}} \sim (250 \mathrm{nm})^{-3}$, and the smallest confinement scale at which nontrivial quasicrystals are stable is $d_z \sim 250$ nm. Thus, the three-body interaction scale is at most $10^{-5}$ as large as the two-body interaction scale, which governs crystallization. It is also much smaller than the temperature (which is at best a tenth of the two-body interaction scale), or the energy scale associated with the harmonic confinement of the system, which breaks translational symmetry and gaps out all the translational modes including the phonons. Consequently, these additional modes in our system have just as legitimate a claim as phonons (or conventional phasons) to be described as gapless.

That said, we agree that the counting of Ref.~\cite{lifshitz-comment} provides a more general lower bound, as it enumerates the \emph{minimal} number of Goldstone modes consistent with being a quantum quasicrystal. Although the additional modes resemble phasons in being continuous rearrangements, they are not consequences of quasiperiodicity (indeed, as we remark~\cite{gmd}, they also occur in crystals), and are perhaps better described as pseudo-phasons.

Finally, we address the observation~\cite{lifshitz-comment} that odd powers of the order parameter can also be forbidden in classical quasicrystals by imposing an additional $Z_2$ symmetry. While this is formally true, there is no obvious \emph{physical} reason that such a symmetry should exist in classical quasicrystals, where the order parameters are Fourier components of the density. In the quantum case, by contrast, the restriction to even-order terms follows because the order parameters are the condensing Bose fields. For a similar situation to occur classically, it seems that crystalline order would either have to be intertwined with some other order (e.g., magnetism~\cite{jagannathan}) or have to involve a more complicated order parameter than the density modulations~\cite{marchenko}. The former case is similar to that of a quantum quasicrystal (as we remark in Ref.~\cite{gmd}), and we do not know of any experimental instances of the latter.

The authors are indebted to R.M. Wilson for helpful discussions.

\end{document}